\documentclass{ws-ijgmmp}
\usepackage{orcidlink}

\begin{document}

\markboth{Authors' Names}
{Cosmological Bounce Scenario with a Novel Parametrization of Bulk Viscosity}

%%%%%%%%%%%%%%%%%%%%% Publisher's Area please ignore %%%%%%%%%%%%%%%
%
\catchline{}{}{}{}{}
%
%%%%%%%%%%%%%%%%%%%%%%%%%%%%%%%%%%%%%%%%%%%%%%%%%%%%%%%%%%%%%%%%%%%%

\title{Cosmological Bounce Scenario with a Novel Parametrization of Bulk Viscosity}

\author{Rajdeep Mazumdar}
\address{Department of Physics, Dibrugarh University, Dibrugarh, Assam-786004, India\\
rajdeepmazumdar377@gmail.com}

\author{Mrinnoy M. Gohain \orcidlink{0000-0002-1097-2124}}
\address{Department of Physics, Dibrugarh University, Dibrugarh, Assam-786004, India\\
mrinmoygohain19@gmail.com}

\author{Kalyan Bhuyan \orcidlink{0000-0002-8896-7691}}
\address{Department of Physics, Dibrugarh University, Dibrugarh, Assam-786004, India\\
Theoretical Physics Divison, Centre for Atmospheric Studies, Dibrugarh University, Dibrugarh, Assam-786004, India\\
kalyanbhuyan@dibru.ac.in}

\maketitle

\begin{history}
\received{(Day Month Year)}
\revised{(Day Month Year)}
\end{history}

\begin{abstract}
In this work, we have studied how incorporating viscous fluids leads to exact bounce cosmological solutions in general relativity (GR) framework. Specifically, we propose a novel parameterization of bulk viscosity coefficient of the form $\zeta = \zeta_0 (t-t_0)^{-2n} H$, where $\zeta_0$,  $n$ being some positive constants and $t_0$ is the bounce epoch. We investigate how this form of bulk viscosity may assist in explaining the early universe's behaviour, with a particular focus on non-singular bounce scenario by studying the various energy conditions and other related cosmological observables and how the model parameters affect the evolution of the Universe. We demonstrate that the NEC and SEC violation occurs at the bounce point while DEC is satisfied. Finally, we carried out a stability check based on linear order perturbation to the Hubble parameter. We found that the perturbation vanishes asymptotically at later times, which indicates a stable behaviour of the bounce scenario.
\end{abstract}


\section{Introduction}\label{sec1}
Einstein's General Relativity (GR) has been regarded as the foundation of physical laws on a grander scale and is essential to astrophysics and cosmology \cite{ref1}. GR is a mathematically simple geometric theory holding significant physical explanation of various observed phenomena including gravitational waves, black holes, quasars and even the dynamics of the Universe as a whole \cite{ref1, ref2}, concerning which different works have been formulated around it ( see Ref. \refcite{ref3} to Ref. \refcite{ref29}). From a cosmological perspective, the basic premise is that the Universe originated with a point of infinite energy density - the Big Bang, an initial space-time singularity that necessitates the understanding of quantum gravity. In addition to the initial singularity problem, conventional Big Bang cosmology raises many other concerns, including the horizon problem, flatness problem, transplanckian problem, and so on. Guth's inflation hypothesis proved effective in resolving the majority of such issues. However, the inflationary scenario is plagued with the singularity problem and the trans-Planckian problem of fluctuations. At the onset of inflation, during which the universe expands almost exponentially, an initial singularity endures causing the inflationary scenario to fail as a complete cosmological theory as a robust and experimentally sound theory of quantum gravity is still lacking. To overcome the aforementioned issue, bouncing cosmologies were proposed \cite{ref30}. This family of theories is purely classical, avoiding the demand for a theory of quantum gravity, thus allowing for the study of the Universe at or before the point of transition to standard cosmological history. The basic idea of bouncing cosmologies is that instead of retracing the expansion back to an infinitesimal point, there was a period when the Universe was at a non-minimal size and finite energy density, beyond which contraction was inconceivable. As an outcome, the Universe does not form a singularity; rather, it emerges from the expansion of the prior contracting phase, thereby avoiding the singularity. The scale factor in the bouncing paradigm decreases to a limited value before expanding, whereas the Hubble parameter grows drastically after reaching zero. The efficiency of bouncing models in solving basic cosmological problems can be visualized through the wedge diagram proposed by Ijjas and Steinhardt \cite{ref31}. Technically, the realization of a bounce requires a violation of the null energy condition. The violation of NEC indicate that with an increase in time, the Hubble parameter $H$ increases and $\dot{H} > 0$. We elaborate below the required properties of bouncing cosmological models \cite{ref32}
\begin{itemize}
    \item At the bouncing epoch, the scale factor contracts to a non-zero finite value, the Hubble parameter vanishes, and the deceleration parameter becomes singular.
    \item The NEC is violated since the Hubble parameter changes sign from the bouncing point; therefore, this phenomenon is ruled out in the context of General Relativity (GR).
    \item The scale factor slope increases after the bounce, while the Hubble parameter remains negative during matter contraction and positive during matter expansion.
\end{itemize} 

Several works have been carried out in the literature based on different frameworks like string-inspired gravity, quintom matter, $f(T)$ gravity, scalar fields (see Ref. \refcite{ref33,ref34,ref35,ref36}, and relevant references therein and Ref. \refcite{ref37} for a review). Various bouncing cosmologies have been formulated to date namely symmetric bounce, super bounce, oscillatory and matter bounce. The symmetric bounce was first considered by Cai et al \cite{ref35} to produce a non-singular bouncing cosmology following an ekpyrotic contraction phase. Nevertheless, this bounce suffers from problems with primordial modes not accessing the Hubble horizon until it is combined with other cosmological behaviours \cite{ref38,ref39,ref40}. Superbounce cosmologies, first proposed by M. Koehn \cite{ref41}, are used to build a universe in which there is no singularity and the universe collapses and resurrects through a Big Bang \cite{ref42}. The power-law scale factor describes this kind of cosmology. Note that a Hubble parameter that varies in signature before and after the bounce but becomes unique at the bounce point characterises the super bounce. An oscillatory scale factor provides oscillatory bouncing cosmologies. The behaviour of a cyclic universe, which views the cosmos as a continuous series of contractions and expansions \cite{ref43,ref44,ref45,ref46}, is represented by such models \cite{ref37}. The matter bounce cosmology that derives from the loop quantum cosmology (LQC) has been studied in the early cosmos and has demonstrated the capacity to create a power spectrum that is either scale-invariant or nearly scale-invariant \cite{ref48,ref49,ref50}. 

As far as the total matter content of the Universe is concerned, viscous fluids 
that impart dissipative effects are pivotal and more generic as a cosmological fluid and that perfect fluids are merely approximations of imperfect fluids. That is why, it is quite reasonable to take it into consideration a physical scenario. The viscous fluids are quantified by viscosity coefficients and are widely used in cosmological models. From a hydrodynamical standpoint, when a system loses thermal equilibrium, an effective pressure is generated to re-establish thermal stability.
If the Universe's total matter composition displays bulk viscosity, then it leads to an effective pressure that can be of major importance in modelling realistic cosmological models. Shear viscosity and bulk viscosity are the two main types of viscosity of cosmological fluids. Fluid velocity gradients and shear viscosity are related to each other. Conversely, bulk viscosity introduces damping which is linked to volumetric straining. Since the cosmic fluid is spatially isotropic, the effects of shear viscosity on cosmic dynamics are negligible. However, bulk viscosity may have a major impact on cosmic history \cite{ref51,ref52,ref53}. Some studies suggested that inflation could be caused by bulk viscous fluids as well as this may leaf to the late time acceleration of the  Universe. These fluids have also been identified as potential candidates for dark matter, dark energy and unified scenarios (see Ref. \refcite{ref54,ref55,ref56} along with Ref. \refcite{ref53,ref56_a,ref56_c,ref56_d} for a review). In this paper, our objective is to investigate how the impact of bulk viscosity coefficient $\zeta$ in cosmic pressure influences the universe's development, particularly in the arena of bouncing cosmology. The incorporation of viscous effects in cosmic fluids can lead to a more relaxed description of the fluid's ideal characteristics and a reduction in total pressure, which may lead to interesting dynamics of the early Universe cosmology.

The work is presented as follows: In section \ref{sec2} we set up the necessary Friedmann equations relevant to the model. In section \ref{sec3} we shall obtain the exact solutions to the Friedmann equations for a model with a viscosity coefficient $\zeta = \zeta_0 (t - t_0)^{-2n} H$. In section \ref{sec4} we study the evolution of physical cosmological observables like energy density and deceleration parameter for the model, along with the concerning energy conditions to validate the model. In section \ref{sec5} we investigate the stability of the model using perturbation analysis. In section \ref{sec6}, we conclude the study with our findings and future perspectives.
\section{Framework}\label{sec2}
The model's basic equations are derived from Einstein's field equations of general relativity, which relate space-time geometry to the universe's energy and matter content, given by:
\begin{equation}
    G_{\mu\nu} = 8 \pi G T_{\mu\nu} \label{eq:1}
\end{equation}
Where, $G_{\mu\nu}$ is the Einstein curvature tensor representing the curvature of space-time, $T_{\mu\nu}$ is the energy-momentum tensor representing the distribution of matter and energy in space-time, $g_{\mu\nu}$ is the metric tensor describing the geometry of space-time, and $G$ is the gravitational constant respectively. The Lambda-CDM model assumes that the universe is homogeneous and isotropic on large scales, meaning it looks the same in all directions. This assumption is supported by observations of the cosmic microwave background radiation and the large-scale structure of the universe.
Here we adopt the spatially flat homogeneous and isotropic FLRW line element given by:
\begin{equation}
    ds^2 = -dt^2 + a(t)^2(dx^2 + dy^2 + dz^2) \label{eq:2}
\end{equation}
where \(a(t)\) is the scale factor. Here, we shall consider the situation where the energy-momentum tensor in the presence of an effective fluid, resulting from the contribution of bulk viscosity to the total matter content, with energy density \(\rho\) and effective pressure \(\tilde{p}\), is defined by:
\begin{equation}
    T_{\mu\nu} = (\rho + \tilde{p})u_\mu u_\nu + \tilde{p} g_{\mu\nu} \label{eq:3}
\end{equation}
where \(u^\mu = (-1,0,0,0)\) is the four-velocity of the fluid satisfying the condition \(u^\mu u_\mu = -1\). The effective fluid considered here is a combination of a perfect fluid and a bulk viscous fluid. The effective pressure is then given by \(\tilde{p} = p - 3H\zeta\), where \(\zeta\) is the bulk viscosity coefficient, \(H\) is the Hubble parameter, and \(\rho\) and \(p\) are the perfect fluid energy density and pressure, respectively, which are connected by the equation of state \(\omega = \rho/p\).
Now, combining Eq. (\ref{eq:1}), Eq. (\ref{eq:2}) and 	Eq. (\ref{eq:3}), the field equations that describe the evolution of the scale factor \(a(t)\) are governed by the following equations:
\begin{equation}
    H^2 = \left(\frac{\dot{a}}{a}\right)^2 = \frac{\rho}{3} \label{eq:4}
\end{equation}
\begin{equation}
    \frac{\ddot{a}}{a} = \dot{H} + H^2 = -\frac{1}{6}(\rho + 3\tilde{p})\label{eq:5}
\end{equation}
Here dot (.) represents one derivative with respect to time. Also, we have set $8\pi G = 1$. These equations describe how the expansion of the universe is influenced by the energy density of matter and radiation, the curvature of space, and the cosmological constant.
\section{Solution}\label{sec3}
\begin{figure}[!h]
\centerline{\includegraphics[scale=0.5]{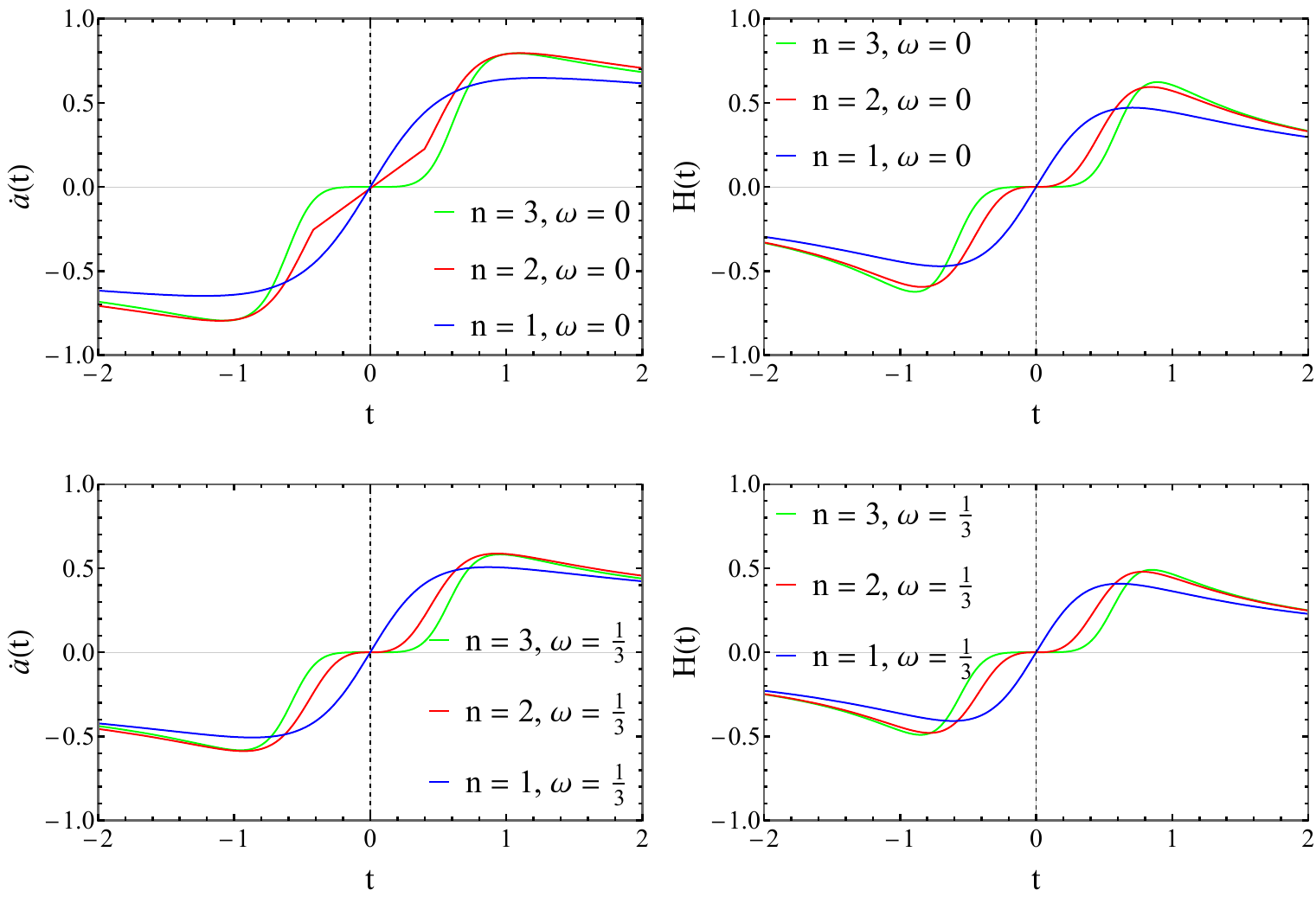}}
\caption{Plot of variation of the first-order time derivative of the scale factor and Hubble parameter with cosmological time for the model for different values of the parameter \(n\). For the plot, we have assumed \(\zeta_0 = 0.5\), \(t_0 = 0\), and \(a_0 = 1\). The units of $t$ and $\dot{a}$ are in Gyr and Gyr$^{-1}$.}
\label{fig1}
\end{figure}

\begin{figure}[!h]
\centerline{\includegraphics[scale=0.5]{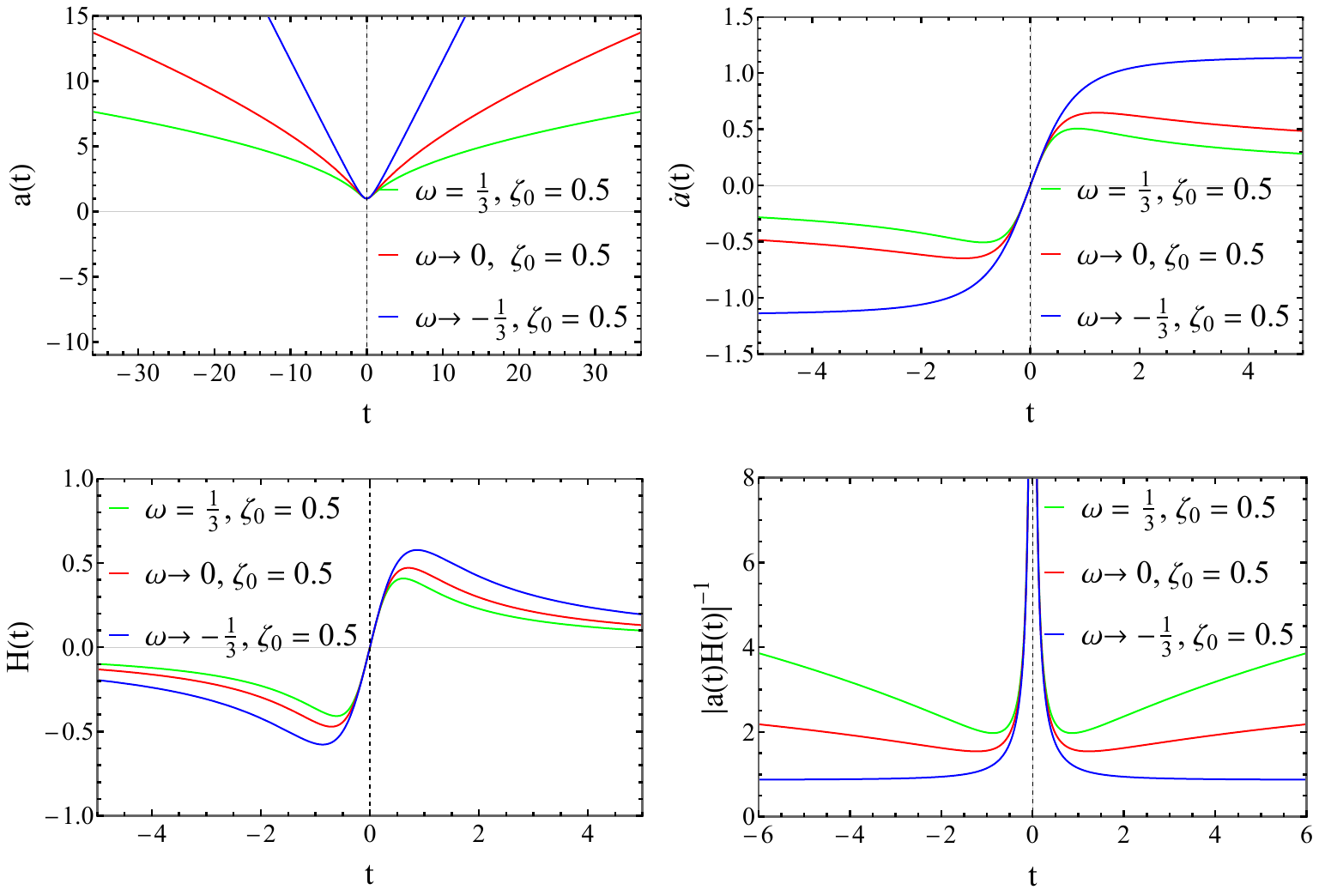}}
\caption{Variation of Scale factor, Hubble parameter, Hubble radius, and the first-order time derivative of the scale factor with cosmological time for model-1. For the plot, we have assumed \(\zeta_0=0.5\), \(n=1\), \(t_0=0\), and \(a_0=1\). The units of $t$ and $H$ are in Gyr and Gyr$^{-1}$.
}
\label{fig2}
\end{figure}

\begin{figure}[!h]
\centerline{\includegraphics[scale=0.5]{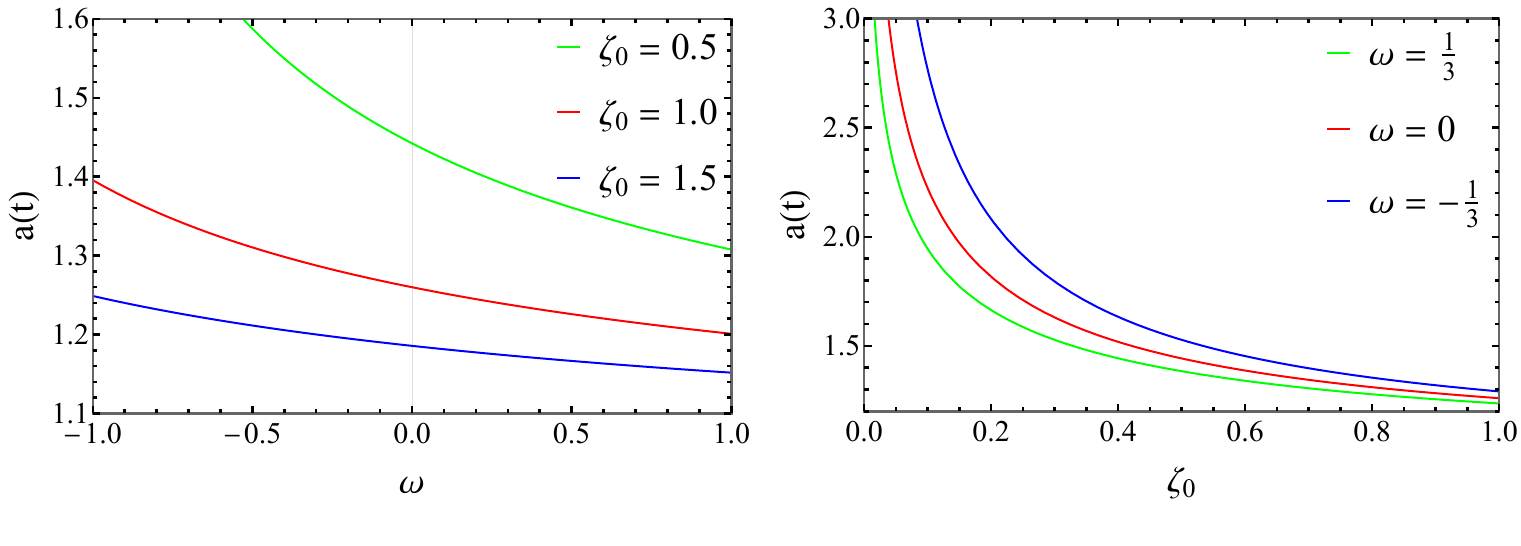}}
\caption{Plot of variation of the Scale factor concerning different values of the model parameter \(\omega\) and \(\zeta_0\). For the plot, we have assumed \(t = 1\), \(n = 1\), \(t_0 = 0\), and \(a_0 = 1\).}
\label{fig3}
\end{figure}

\begin{figure}[!h]
\centerline{\includegraphics[scale=0.5]{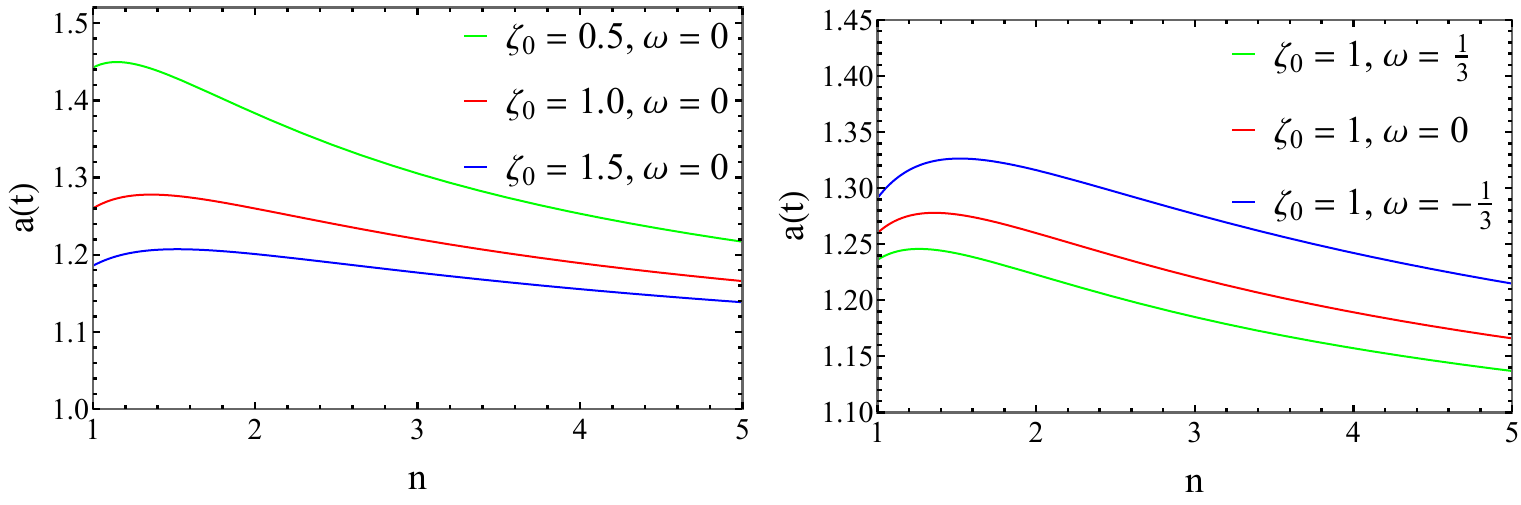}}
\caption{Plot of variation of the Scale factor with respect to different values of the model parameter \(n\). For the plot, we have assumed \(t = 1\), \(t_0 = 0\), and \(a_0 = 1\).}
\label{fig4}
\end{figure}

In this section, we focus on the possibility of finding the exact solution to the field equations given by Eq. (\ref{eq:4}) and Eq. (\ref{eq:5}) in the presence of a bulk viscous fluid by considering a modified bulk viscosity coefficient. Generally, in cosmological scenarios bulk viscosity coefficients are taken to be functions of energy density (and hence Hubble parameter) or higher powers of it, which in retrospect can functions of velocity or acceleration dependent terms \cite{ref52, ref53}. In the light of this possibility, let us propose a new parameterization of the bulk viscosity coefficient:
%\cite{ref34,ref35,ref36,ref37,ref38,ref39,ref40,ref41,ref42}. 
\begin{equation}
    \zeta = \zeta_0 (t - t_0)^{-2n} H \label{eq:6}
\end{equation}
where \(\zeta_0\), \(t_0\), and \(n\) are some positive constant parameters and \(H\) is the Hubble parameter. Here, \(n\) is dimensionless, \(t_0\) has the dimension of time, and similarly following which \(\zeta_0\) has dimensions \([\text{M}]^1 [\text{L}]^{-1} [\text{T}]^{2n}\) to maintain the dimensional correctness of the Eq. \ref{eq:6}.\footnote{If one chooses geometric units with $cm$ as the basic unit, the dimension of the pressure will be  1/cm$^4$, which implies that the dimension of $\zeta$ is 1/cm$^3$. Then, according to Eq.\eqref{eq:6}, the dimension of $\zeta_0$ becomes $[\zeta_0]= cm^{(2n-2)}$. For $n = 1$, $\zeta_0$ becomes dimensionless. The units of $\zeta_0$ for different values of $n$ are different, but nevertheless consistent with the overall dimension of $\zeta$ as seen from Eq. \eqref{eq:6}}From a somewhat related perspective, Brevik and Timoshkin worked out holographic bounce by considering a bulk viscous fluid \cite{ref56_b}. Then using Eq. (\ref{eq:4}) and Eq. (\ref{eq:6}) , we can rewrite Eq. \eqref{eq:5} into the following form:
\begin{equation}
    \dot{H} = (\alpha (t - t_0)^{-2n} - \beta) H^2 \label{eq:7}
\end{equation}
where \(\alpha = \frac{3}{2} \zeta_0\) and \(\beta = \frac{3}{2} (1 + \omega)\) respectively. The Hubble parameter is obtained by solving Eq. \eqref{eq:7} as:
\begin{equation}
    H(t) = \frac{(-1 + 2n)(t - t_0)^{2n}}{-t_0\alpha + t(\alpha + (-1 + 2n)(t - t_0)^{2n}\beta) - (-1 + 2n)(t - t_0)^{2n}C_1} \label{eq:8}
\end{equation}
where \(C_1\) is an integration constant. Now, as \(H = \frac{\dot{a}}{a}\), Eq. (\ref{eq:8}) can further be integrated to obtain the scale factor as:
\begin{equation}
    a(t) = C_2 \left(n(\alpha + (-1 + 2n)(t - t_0)^{2n}\beta)\right)^{\frac{1}{2n\beta}}, \label{eq:9}
\end{equation}
where we have assumed the integration constant \(C_1 = t_0\beta\) to obtain the solution in a closed form. \(C_2\) is again an integration constant, and subject to the condition that \(a(t) \to a_0\) when \(t \to t_0\), where $a_0$ is the scale factor and $t_0$ is the bounce epoch, we get \(C_2 = a_0 \left(n\alpha\right)^{-\frac{1}{2n\beta}}\). Thus, we have
\begin{equation}
    a(t) = a_0 \left(n\alpha\right)^{-\frac{1}{2n\beta}} \left(n(\alpha + (-1 + 2n)(t - t_0)^{2n}\beta)\right)^{\frac{1}{2n\beta}} \label{eq:10}
\end{equation}
As \(H = \frac{\dot{a}}{a}\)  we can rewrite the expression for the Hubble parameter as:
\begin{equation}
    H(t) = \frac{(2 n-1) (t-t_0)^{2 n-1}}{\alpha +\beta  (2 n-1) (t-t_0)^{2 n}}\label{eq:10.1}
\end{equation}
From Fig. \ref{fig1} and Fig. \ref{fig2}, we can see for different possible values of \(n\) the scale factor increases (\(\dot{a} > 0\)) during the expanding phase (positive time zone) and decreases (\(\dot{a} < 0\)) during the contracting phase (negative time zone). It yields \(\dot{a} = 0\) at \(t = t_0\). Similarly, we see the Hubble parameter makes a transition from a negative value (Contracting Universe) to a positive value (Expanding Universe) at \(t = t_0\), and the Hubble radius is seen to diverge at \(t = t_0\) respectively. It is evident that the model under consideration represents a bouncing cosmology where the bouncing takes place at the cosmological time \(t = t_0\), as it exhibits the preliminary behaviour for a bouncing cosmology.
To understand explicitly the dependence of the scale factor obtained from the model represented by the Eq. (\ref{eq:10}) on the model parameters \(\omega\), \(\zeta_0\) and  \(n\), we have numerically studied the variation of the scale factor with the parameters at cosmological time \(t = 1\) for \(a_0 = 1, t_0 = 0\) as shown in Fig. \ref{fig3} and Fig. \ref{fig4}. From this it is relevant that the scale factor decreases with the increasing value of the parameter \(\zeta_0\) and the scale factor increases with the decrease in the value of the parameter \(\omega\). Along with that, it is observed that the dependence of the scale factor on the parameter \(n\) is also significant.
\section{Evolution of Cosmological observables and Energy Conditions}\label{sec4}
In this section, let us study the evolution of some relevant cosmological observables like the energy density and the deceleration parameter, along with the concerning energy conditions represented by the model to test its validity at the bounce point. 
\subsection{Energy Density}
\begin{figure}[!h]
\centerline{\includegraphics[scale=0.5]{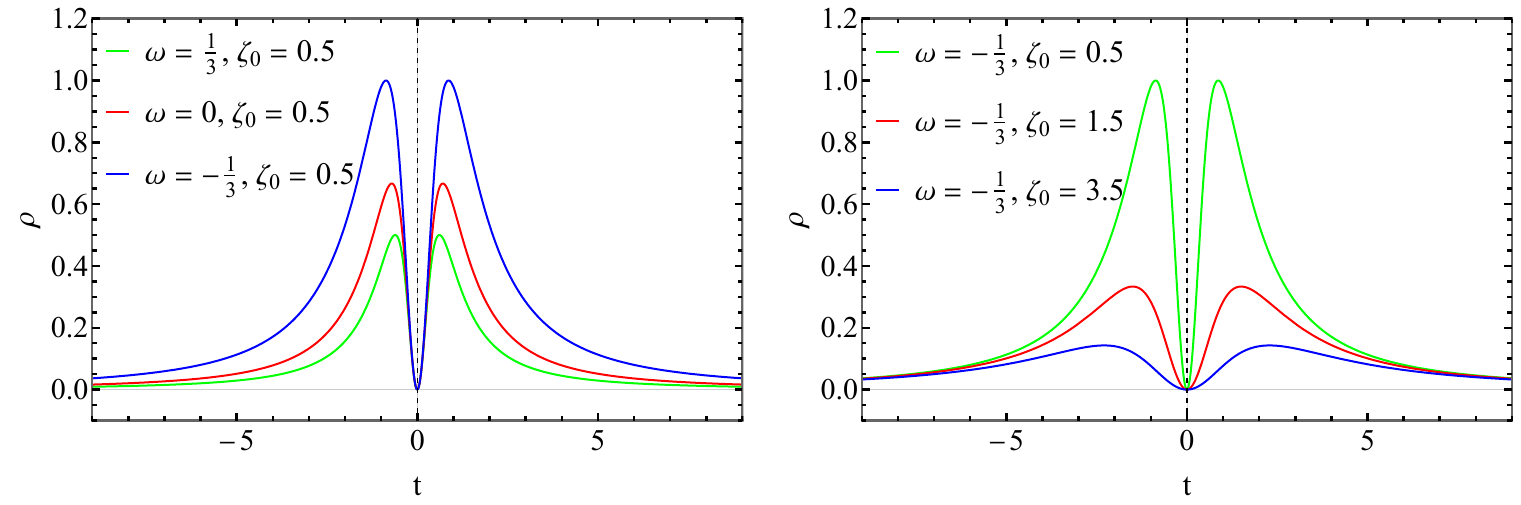}}
\caption{Evolution of the energy density with cosmological time \( t \) for the Model. For the plot, we have taken \( n = 1 \), \( t_0 = 0 \), and \( a_0 = 1 \). The unit of $t$ is in Gyr.}
\label{fig5}
\end{figure}

Using the Eq. (\ref{eq:4}) and Eq. (\ref{eq:10}), we can obtain the energy density for the model under consideration as:
\begin{equation}
    \rho = \frac{4}{3 (t - t_0)^2 \left(1 + \left(\frac{(t - t_0)^{-2n} \zeta_0}{-1 + 2n} + \omega\right)^2\right)} \label{eq:11}
\end{equation}
Fig. \ref{fig5} shows the evolution of the energy density \(\rho\) with the cosmological time \(t\) for different combinations of model parameters. It can be seen that during the contracting phase (negative time zone), the energy density initially increases with the increasing cosmological time until it attains a maximum value, after which the energy density starts decreasing with the increasing cosmological time, finally becoming zero at the bouncing epochs. 
Similarly, for the expanding phase (positive time zone) we can see that the energy density starts from zero at the bouncing epochs, then it starts increasing with the increasing cosmological time until it attains a maximum value, after which the energy density starts decreasing with the increasing cosmological time. Nevertheless, the energy density maintains a non-negative value throughout all the time which is expected and required. The violation of NEC is important at the bouncing epoch, which constitutes the depletion of the energy density. Hence, the depletion of energy density at the bouncing point in Fig. \ref{fig5} indicates the possibility of the violation of NEC. The results show a resemblance to the ones obtained in different works \cite{ref57,ref58,ref59}.

\subsection{Deceleration Parameter}
In cosmology, the deceleration parameter is a quantity used to express the rate at which the universe is slowing down its expansion. It measures how the universe's pace of expansion has changed throughout time. The deceleration parameter \( q \) is defined as:
\begin{equation}
    q = -\frac{\ddot{a}a}{\dot{a}^2} \label{eq:12}
\end{equation}
For \( q > 0 \), we have a deceleration, and for \( q < 0 \), we have accelerating universe scenarios. According to the work of Singh \& Bishi \cite{ref60} cosmological models are divided into the following categories based on their time dependence on the Hubble parameter and deceleration parameter, as shown in Table \ref{tab:1} respectively. From the classification, cases (I), (II), and (III) are possible, as in the present scenario the Universe is expanding. Depending on that, there are different types of expansion exhibited for our Universe, as given in Table \ref{tab:2} respectively \cite{ref60}. Using Eq. (\ref{eq:10}) and Eq. (\ref{eq:12}), we find the deceleration parameter \( q \) for our model as:
\begin{equation}
    q = \frac{1}{2} \left(1 - 3(t - t_0)^{-2n} \zeta_0 + 3\omega\right) \label{eq:13}
\end{equation}

\begin{table}[tbh]
\tbl{Cosmological scenarios classification based on Hubble and deceleration parameter.}
{\begin{tabular}{@{}cccc@{}} \toprule
\textbf{Case} & \textbf{Condition} & \textbf{Universe Scenario} \\
\colrule
I    & \(H > 0, q > 0\)     & Expanding and Decelerating\\
II   & \(H > 0, q < 0\)     & Expanding and Accelerating\\
III  & \(H > 0, q = 0\)     & Expanding and Zero Deceleration/Constant Expansion\\
IV   & \(H < 0, q > 0\)     & Contracting and Decelerating\\
V    & \(H < 0, q < 0\)     & Contracting and Accelerating\\
VI   & \(H < 0, q = 0\)     & Contracting and Zero Deceleration/Constant Expansion\\
VII  & \(H = 0, q = 0\)     & Static\\
\botrule
\end{tabular} \label{tab:1}}
\end{table}
\begin{table*}[tbh]
\tbl{Cosmological Scenarios for an expanding universe with different deceleration parameter values.}
{\begin{tabular}{@{}ccc@{}} \toprule
\textbf{Case} & \textbf{Condition} & \textbf{Universe Scenario}\\
\colrule
A & \(q < -1\) & Super Exponential Expansion\\
B & \(-1 \leq q < 0\) & Exponential Expansion (known as de-Sitter expansion)\\
C & \(q = 0\) & Expansion with Constant Rate\\
D & \(-1 < q < 1\) & Accelerating Power Expansion\\
E & \(q > 0\) & Decelerating Expansion\\
\botrule
\end{tabular} \label{tab:2}}
\end{table*}
\begin{figure}[!h]
\centerline{\includegraphics[scale=0.5]{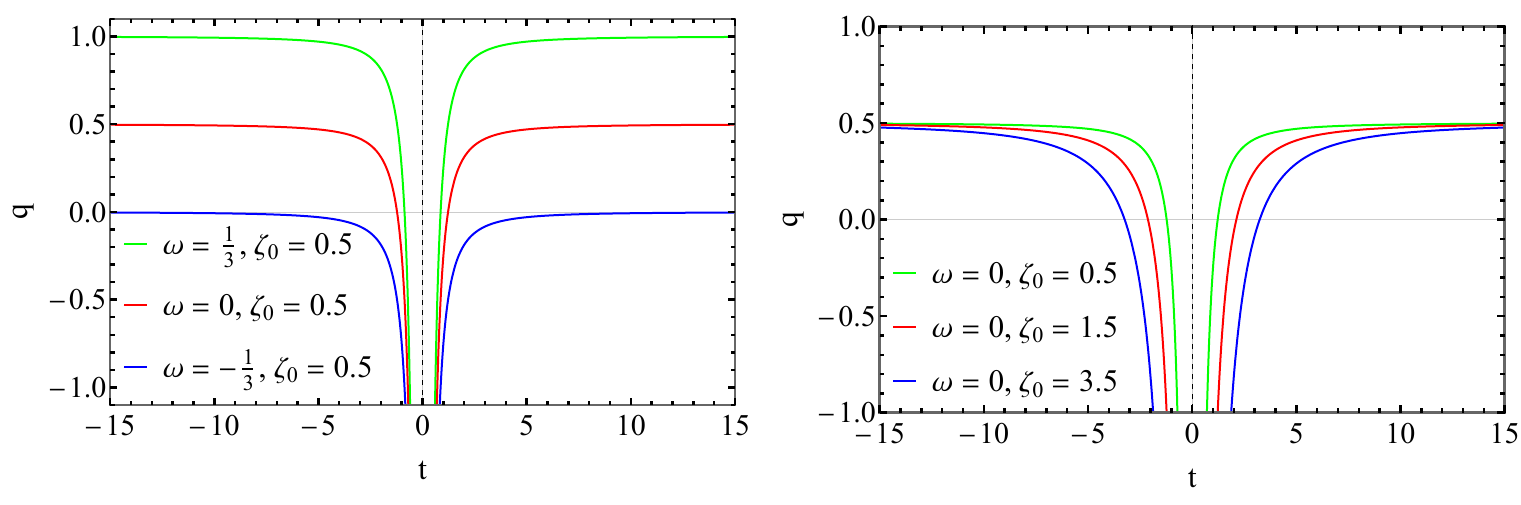}}
\caption{Evolution of the deceleration parameter with cosmological time \( t \) for the Model. For the plot, we have taken \( n = 1 \), \( t_0 = 0 \), and \( a_0 = 1 \). The unit of $t$ is in Gyr.}
\label{fig6}
\end{figure}
In Fig. \ref{fig6}, the behaviour of the deceleration parameter at the bouncing point \( t = 0 \) for different combinations of $(\omega, \zeta_0)$ are shown. Evidently, for some combinations $(\omega, \zeta_0) : (1/3, 0.5), (0, 0.5), (0, 1.5), (0, 3.5)$ it seems to exhibit a symmetrical behaviour at the bouncing point and evolves with negative values after the bounce, and transits into the $q > 0$ zone, representing decelerated matter dominated phase. However $(\omega, \zeta_0) = (-1/3, 0.5)$ does not show the transition from the accelerated to the decelerated phase.
\subsection{Energy Conditions}
The Friedmann equation provides a set of energy conditions that may be used to predict the cosmic acceleration in current cosmology. These energy conditions play a crucial role in GR by proving the theorems for the existence of black holes and space-time singularity \cite{ref61}. Hence, to verify the model's viability in the setting of cosmic acceleration, we shall examine the well-known energy requirements in this section. For the content of the universe expressed as a viscous fluid, many types of energy conditions are provided, including weak energy conditions (WEC), null energy conditions (NEC), dominant energy conditions (DEC), and strong energy conditions (SEC), which are given as \cite{ref62}:
\begin{itemize}
    \item Null Energy Condition (NEC) $\iff \rho + \tilde{p} \geq 0$
    \item Strong Energy Condition (SEC) $\iff \rho + 3\tilde{p} \geq 0$
    \item Dominant Energy Condition (DEC) $\iff \rho - \tilde{p} \geq 0$
    \item Weak Energy Condition (WEC) $\iff \rho \geq 0, \rho + \tilde{p} \geq 0$
\end{itemize}
For our model, we find the effective pressure \( \tilde{p} \) as:
\begin{align}
    \tilde{p} &= -\frac{4(1-2n)^2 (t - t_0)^{-2+2n} (\zeta_0 - (t - t_0)^{2n} \omega)}{3\left(\zeta_0 - (t - t_0)^{2n} (1+\omega) + 2n (t - t_0)^{2n} (1+\omega)\right)^2} \label{eq:15}
\end{align}
\begin{figure}[!h]
\centerline{\includegraphics[scale=0.55]{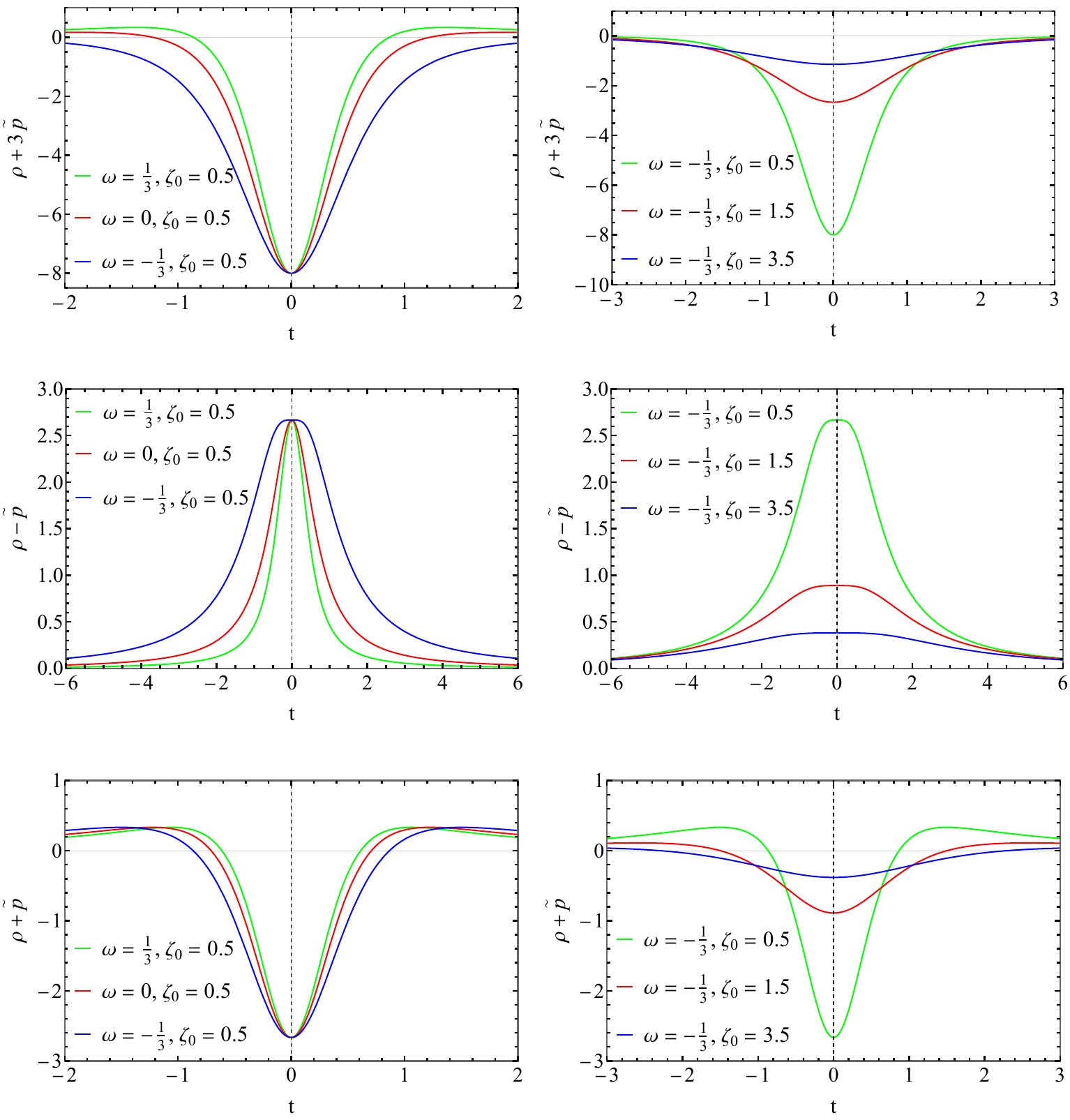}}
\caption{Plot for the variation of the energy conditions versus cosmological time for the Model. For the plot, we have taken \( n = 1 \) and \( t_0 = 0 \). The unit of $t$ is in Gyr.}
\label{fig7}
\end{figure}
Now using Eq. \ref{eq:12} and Eq. \ref{eq:15}, we evaluate $\rho + \tilde{p}$, $\rho + 3 \tilde{p}$ and $\rho - \tilde{p}$. Because of the highly complicated form of the equations, we have refrained from expressing its mathematical form. Instead, we present a graphical evolution of them with respect to cosmological time as shown in Fig. \ref{fig7}. The violation of NEC constitutes the depletion of the energy density as the universe expands. The violation of SEC constitutes the acceleration of the universe. The realization of a bounce requires a violation of the NEC at the bounce point. The SEC must be broken to describe a universe that is governed by negative pressure \cite{ref63}. According to the most recent data on the expanding universe, the SEC must be broken on an astronomical scale \cite{ref64,ref65}. The behaviour of SEC shows how quickly the Universe is expanding. From Fig. \ref{fig7} it is evident that at the bounce point, the NEC and SEC are both broken. As the Universe expands, the bouncing scenario has to incorporate a basic violation of the NEC, illustrating the depletion of the energy density. The appearance of dark energy-like behaviour at later times is caused by the SEC violation, which satisfies the above-indicated requirement for the model to conform with current observations of the expanding Universe. The fact that the NEC and SEC are non-singular at the bouncing point solves the singularity problem of the early description and provides evidence for a non-singular bouncing universe. We also see the violation of WEC, but the DEC is expected to be satisfied in the context of a perfect fluid and the same is also observed in the analysis.
\section{Stability Analysis}\label{sec5}
In this section, we are interested in investigating the model's stability through a method of perturbation analysis of the Hubble parameter. To do so, we have presumed a linear order perturbation of the Hubble parameter given as
\begin{equation}
    H(t)^*=H(t) \left(\delta(t)_m +1\right),  \label{eq:26}
\end{equation}
where \(H(t)^*\) is the perturbed Hubble parameter and \(\delta(t)_m\) is the perturbation term. It is expected that for a stable scenario, the perturbations must die out with increasing time or at least remain finite.
As we know, the conservation equation for an effective fluid that is a combination of a non-interacting perfect fluid and a bulk viscous fluid can be given as \cite{ref67}
\begin{equation}   
 \dot{\rho} + 3 H (\rho + 3 \tilde{p}) = 0 \label{eq:27}
\end{equation}
Now, using the Eq. (\ref{eq:26}), Eq. (\ref{eq:10.1}), Eq. (\ref{eq:4}) and \(\tilde{p} = p - 3H\zeta\) in Eq. (\ref{eq:27}) we have
\begin{equation}
\begin{aligned}
&8 (2 n-1)^3 \left(\delta _m(t)+1\right){}^2  \left(t-t_0\right){}^{4 n-3} \left(-2 \zeta _0-3 \zeta _0 \delta _m(t)+(3 \omega +1) \delta _m(t) \left(t-t_0\right){}^{2 n} \right. \\& \left.\hspace{8cm}+2 \omega  \left(t-t_0\right){}^{2 n}\right)=0
\end{aligned} \label{eq:28}
\end{equation}
Solving Eq. (\ref{eq:28}) we obtain the perturbation term \(\delta_m(t)\) as
\begin{equation}   
 \delta_m(t) =\frac{2 (\zeta_0 -\omega(t-t_0)^{2 n})}{(3 \omega +1) (t-t_0)^{2 n}-3 \zeta_0}\label{eq:30}
\end{equation}
\begin{figure}[!h]
\centerline{\includegraphics[scale=0.5]{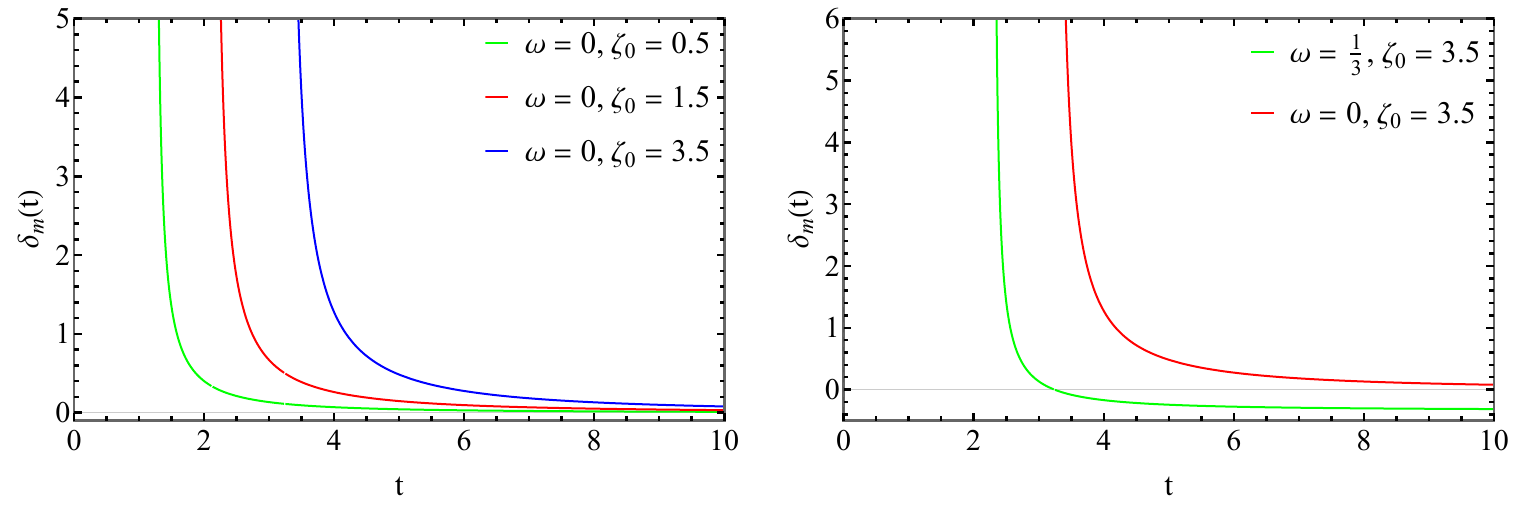}}
\caption{Evolution of the perturbation term \(\delta_m(t)\) with cosmological time for the model. For the plot we have taken \( n = 1 \) and \( t_0 = 0 \). The unit of $t$ is in Gyr.}
\label{figsta}
\end{figure}
Fig. \ref{figsta} displays the evolution of the perturbation with cosmic time for different combinations of $(\omega, \zeta_0)$. The perturbation term \(\delta_m(t)\) seems to evolve eventually towards zero with increasing time. The perturbations rapidly decrease away from the bouncing region, guaranteeing stability at late times.
\section{Conclusion}\label{sec6}
Notably, the initial singularity problem has been the major problem in early Universe cosmology. To address this issue, ``bouncing cosmology" has been proposed as an appealing approach. One may achieve bouncing cosmology by varying curvature or matter components in GR. This paper addresses the non-singular bouncing universe scenario in GR, with a specific emphasis on a viscous fluid at cosmological scales. This may be relevant because it may result in the violation of energy conditions, particularly the NEC, which is a crucial requirement at the point of bounce. In this work, we proposed a novel parameterization of the bulk viscosity coefficient given as $\zeta = \zeta_0 (t - t_0)^{-2n} H$, where where \(\zeta_0\), \(t_0\), and \(n\) are some positive constant parameters and \(H\) is the Hubble parameter.

In section \ref{sec3}, we obtained the exact solutions to the Friedmann equations for a model with the proposed form of viscosity coefficient. The obtained solutions of the model seem to meet some of the basic requirements of a bouncing universe scenario like the Hubble parameter vanishes at the bouncing epoch and the slope of the scale factor slope is negative during the contraction phase and positive during the expansion phase. We studied the physical parameters of our model like energy density and deceleration parameter. It exhibits symmetrical behaviour at the bouncing point and evolves with negative values after the bounce, thereafter transits into the decelerated matter-dominated phase. To see whether our model adheres to the other required criteria like violation of energy conditions at the bounce epoch, we studied the NEC, SEC and DEC evolution in section \ref{sec4}. A non-singular bouncing universe require violations of the NEC and SEC, which we observed at the bouncing point. Furthermore, it turns out that during the bounce, the DEC is satisfied. We examine the model's stability in section \ref{sec5} by performing a perturbation analysis on the Hubble parameter. The perturbation term is observed to asymptotically vanish with increasing time, indicating that the perturbations eventually die out rendering the model stable at a later time.

Although our model meets the requirements for bouncing cosmology, it is important to highlight that other important observable parameters like tensor to scalar ratio, scalar spectral index etc. are not addressed in this work because it is beyond its scope of inquiry. We desire to look into this possibility in future work. An analysis pertaining to the effect of bulk viscosity on the observable parameters might be fascinating.

%\begin{thebibliography}{000} %for 3 digits
%\begin{thebibliography}{00}  %for 2 digits

\end{document}